\documentclass[prc,twocolumn,showpacs,showkeys]{revtex4}
\RequirePackage{amsmath2000}

\usepackage{amsmath2000}
\usepackage[dvips]{graphicx}
\usepackage{dcolumn}
\usepackage{times}

\newcommand{\bea}{\begin{eqnarray}}
\newcommand{\eea}{\end{eqnarray}}
\newcommand{\be}{\begin{equation}}
\newcommand{\ee}{\end{equation}}
\newcommand{\bwt}{\begin{widetext}}
\newcommand{\ewt}{\end{widetext}}

\renewcommand{\vec}[1]{\boldsymbol{#1}}
\newcommand{\im}{{\mathrm{i}}}
\newcommand{\del}[1]{\vec{\nabla\!}_{\vec{#1}}}
\newcommand{\delsq}[1]{\nabla_{\!#1}^2}
\newcommand{\expl}[1]{{\mathrm{exp}}\left[#1\right]}
\renewcommand{\exp}[1]{{\mathrm{e}}^{#1}}
\newcommand{\dif}{{\mathrm{d}}}
\newcommand{\deriv}[1]{\frac{\dif}{\dif #1}}
\newcommand{\dderiv}[1]{\frac{\dif^2}{\dif #1^2}}
\newcommand{\derivof}[2]{\frac{\dif #2}{\dif #1}}

\newcommand{\bra}[1]{\langle#1|}
\newcommand{\ket}[1]{|#1\rangle}
\newcommand{\braket}[2]{\langle#1|#2\rangle}
\newcommand{\conj}[1]{\left(#1\right)^*}
\newcommand{\commutator}[2]{\left[#1,#2\right]}

\newcommand{\ad}{\mathrm{ad}}
\newcommand{\eik}{\mathrm{eik}}

\newcommand{\Psip}{\Psi_{\vec{K}}^{(+)}}

\newcommand{\Psiadp}{\Psi_{\vec{K}}^{\ad(+)}}
\newcommand{\psiadp}{\psi_{\vec{K}}^{\ad(+)}}
\newcommand{\Psiadm}{\Psi_{\vec{K'}}^{\ad(-)}}
\newcommand{\psiadm}{\psi_{\vec{K'}}^{\ad(-)}}

\newcommand{\chip}{\chi_{\vec{K}}^{(+)}}
\newcommand{\chim}{\chi_{\vec{K'}}^{(-)}}
\newcommand{\chil}{\chi_\ell^{(+)}}

\newcommand{\chibara}[1]{\bar{X}_{#1}^{(+)}{}}
\newcommand{\chibarl}{\bar{X}_{K\ell\ell'}^{(+)}{}}

\newcommand{\greensad}{G_{\ad}^{(+)}}
\newcommand{\hvcpe}{H_{vc}+\varepsilon_0}
\newcommand{\hme}{H-\varepsilon}
\newcommand{\tooad}{T_{00}^{\ad}}
\newcommand{\dtooad}{\Delta\tooad}

\newcommand{\rcore}{\vec{R}-\alpha\vec{r}}
\newcommand{\rvale}{\vec{R}-\beta\vec{r}}

\newcommand{\Rr}{\vec{R},\vec{r}}

\newcommand{\eiark}{\exp{-\im\alpha\vec{r}\cdot(\vec{K_R}-\vec{K})}}
\newcommand{\eiarkp}{\exp{\im\alpha\vec{r}\cdot(\vec{K_R}-\vec{K'})}}
\newcommand{\kkp}{(\vec{K},\vec{K'})}
\newcommand{\xkkp}{{\mathrm{X}}\kkp}
\newcommand{\khat}{\vec{\hat{K}}}
\newcommand{\kphat}{\vec{\hat{K}'}}
\newcommand{\bzk}[1]{(\vec{b}+#1\khat)}
\newcommand{\bzkp}[1]{(\vec{b'}+#1\kphat)}
\newcommand{\intb}{\int_{0}^{\infty} b\;\dif b \;J_0(Qb)}
\newcommand{\intr}[2]{\int_{#1}^{#2}R^2\dif R}
\newcommand{\comkv}{\commutator{\vec{K_R}}{V_{cT}}}
\newcommand{\xexact}{\tilde{X}_\ell}
\newcommand{\asym}{\underset{R\to\infty}{\sim}}
\newcommand{\cona}{\frac{\hbar^2}{2\mu}}

\newcommand{\spharm}[2]{Y_{#1}(\vec{\hat{#2}})}
\newcommand{\spharmc}[2]{Y_{#1}^{*}(\vec{\hat{#2}})}

\begin{document}

\title{Non-adiabatic corrections to elastic scattering of halo nuclei}

\author{N.~C.~Summers}
 \email{n.summers@surrey.ac.uk}
\author{J.~S.~Al-Khalili}
\author{R.~C.~Johnson}
\affiliation{Department of Physics, School of Physics and Chemistry, University 
of Surrey, Guildford, Surrey, GU2 7XH, United Kingdom}

\date{May 1, 2002}

\begin{abstract}
We derive the formalism for the leading order corrections to the adiabatic 
approximation to the scattering of composite projectiles.
Assuming a two-body projectile of core plus loosely-bound valence particle and a 
model (the core recoil model) in which the interaction of the valence particle 
and the target can be neglected, we derive the non-adiabatic correction terms 
both exactly, using a partial wave analysis, and using the eikonal 
approximation.
Along with the expected energy dependence of the corrections, there is also a 
strong dependence on the valence-to-core mass ratio and on the strength of the 
imaginary potential for the core-target interaction, which relates to absorption 
of the core in its scattering by the target.
The strength and diffuseness of the core-target potential also determine the 
size of the corrections.
The first order non-adiabatic corrections were found to be smaller than 
qualitative estimates would expect.
The large absorption associated with the core-target interaction in such halo 
nuclei as $^{11}$Be kills off most of the non-adiabatic corrections.
We give an improved estimate for the range of validity of the adiabatic 
approximation when the valence-target interaction is neglected, which includes 
the effect of core absorption. 
Some consideration was given to the validity of the eikonal approximation in our 
calculations.
\end{abstract}



\maketitle

\section{Introduction}
Halo nuclei are often described by few-body models which recognise selected 
internal degrees of freedom of the nucleus.
Few-body reaction models assume the projectile nucleus consists of core and 
valence clusters, which interact with the target while also interacting with 
each other.
Using this few-body approach, excitations of the projectile to both bound and 
continuum states may be included within the reaction model; this breakup is 
significant for weakly bound systems, such as halo nuclei.
Further approximations are usually necessary to handle the breakup continuum.

One such approach which includes strong couplings to the continuum is the 
adiabatic approximation \cite{adia}, in which the breakup continuum is collapsed 
onto a single channel which is degenerate with the ground state of the 
projectile.
This is often referred to in the Coulomb excitation literature as the sudden 
approximation.
Another approach to solving the few-body problem, without requiring the use of 
the adiabatic approximation, is the Coupled Discretised Continuum Channels 
(CDCC) method.
Here, convergence of the observables may be obtained in most cases without the 
need for further approximations.
CDCC methods, in which the breakup continuum is truncated and discretised, have 
been effectively applied to reactions involving two-body projectiles, such as 
deuterons and single neutron halo nuclei.
However, the adiabatic approach provides a simpler calculation scheme and as 
yet, is the only method for dealing with multi-channel effects, including the 
continuum, for three-body projectiles.
Three-body models are frequently used for two-neutron halos such as $^6$He, 
$^{11}$Li, and $^{14}$Be.

A special case of the adiabatic model is when the interaction between the core 
and the target dominates and the interaction between the valence particle and 
the target can be neglected \cite{crm-prl}; we call this the \emph{core recoil 
model}, since the only way the projectile can be excited is through recoil of 
the core in its scattering by the target.
This has been applied to the elastic scattering of halo nuclei, such as 
$^{11}$Be, where the core-to-valence mass ratio is large \cite{crm-prl}.
The core recoil model is also particularly applicable to Coulomb dominated 
process when the valence particle is neutral, and has been applied to the 
Coulomb breakup of high energy deuterons \cite{crm-coul-bu-d,crm-coul-bu+heavy} 
and for one and two neutron halos \cite{crm-coul-bu-1n,crm-coul-bu-2n}.

With the current interest in radioactive nuclear beams operating at energies in 
the region of tens of MeVs, the validity of high energy models, such as the 
adiabatic model, at lower energies is of importance.
Comparisons of CDCC and adiabatic methods for deuteron induced reactions have 
been extensively studied \cite{deuteron-review}.
The adiabatic model was shown to be accurate at much lower energies than one 
would expect from qualitative estimates based on the basic assumption of the 
adiabatic approximation---that the breakup energies of the projectile excited 
during the reaction are small relative to the centre of mass energy of the 
projectile.
A simple estimate of the accuracy of the adiabatic approximation for elastic 
scattering and elastic breakup is given in Ref.~\cite{nac-greece} within the 
core recoil model.
However, this estimate did not take into consideration processes such as core 
absorption, which has been suggested as an important factor in improving the 
accuracy of the adiabatic approximation \cite{deuteron-review}.

The adiabatic approximation underlies many of the microscopic theories for 
nuclear reaction, such as Glauber theory \cite{Glauber}.
Glauber models have been extensively applied to reactions involving deuterons 
and light halo nuclei \cite{Gl-d,4body-coul} and have been useful for extracting 
information on halo sizes due to their diffuse nature 
\cite{halo-radii,halo-radii-l}.
Much work has gone into correcting the eikonal assumptions made in Glauber 
theory \cite{non-eikonal,non-eik-sm}.
Corrected calculations compare well to few-body calculations which make only the 
adiabatic approximation, based on the method of 
Refs.~\cite{adia-method,adia-code,4b-adia}.

The formulae for the leading order non-adiabatic corrections in the core recoil 
model were given first in Ref.~\cite{nac-jpg}.
This paper derived the leading order corrections using the eikonal 
approximation, and discussed the relevance of the terms involved.
Here, we calculate these corrections for $^{11}$Be and $^6$He elastic scattering 
from a $^{12}$C target at 10 MeV/nucleon.
The non-adiabatic corrections were found to be identically zero for a pure point 
Coulomb potential \cite{nac-jpg}.
The cross terms for the case of nuclear+Coulomb interactions are also small for 
light ion reactions \cite{phd-ncs}.
The accuracy of the adiabatic approximation for small angle Coulomb scattering 
was discussed in Ref.~\cite{nac-greece}.
In this paper, all Coulomb effects are ignored.
It is felt that this provides an adequate framework for discussing the validity 
of the adiabatic approximation, although no comparison with experimental data is 
possible.
A full comparison of the adiabatic approximation with experimental data, for the 
elastic scattering of a single neutron halo, including Coulomb effects, has been 
made in Ref.~\cite{crm-prl}.

The format of the paper is as follows.
In Section~\ref{SEC:AD} we discuss the adiabatic approximation in the special 
case of the core recoil model, where the valence-target interaction is 
neglected.
The validity of the adiabatic approximation in the core recoil model is examined 
using qualitative arguments, and comparisons with CDCC calculations are made.
In Section~\ref{SEC:NAC} we rederive the formulae in Ref.~\cite{nac-jpg} using 
an alternative method which expresses the internal Hamiltonian operator for the 
projectile as the product of two operators.
We provide an eikonal derivation as in Ref.~\cite{nac-jpg} and then go on to 
derive the corrections exactly using a partial wave sum.
The eikonal approximation provides useful insights into the nature of the 
non-adiabatic corrections while the exact calculation is used to examine the 
range of validity of the eikonal approximation to the corrections.
This provides some justification for the use of the eikonal approximation in 
more complete calculations which include the valence-target interaction.
This will be reported elsewhere \cite{nac-3bg,phd-ncs}.
In Section~\ref{SEC:RESULTS}, we present calculations of the non-adiabatic 
correction in the core recoil model for $^{11}$Be and $^6$He.
Section~\ref{SEC:NEW-EST} gives an improved estimate for the range of validity 
for the adiabatic approximation.
A summary and concluding remarks are given in Section~\ref{SEC:CONCLUSION}.

\section{Adiabatic approximation \label{SEC:AD}}
Here we describe the projectile using a two-body model.
The position of the valence particle ($v$) relative to the core ($c$) is 
described by the vector $\vec{r}$.
The position vector of the centre of mass of the whole projectile from the 
target is $\vec{R}$.
The few-body Hamiltonian is
\be
H = T_{\vec{R}} + V_{cT}(\vec{R}-\alpha\vec{r}) + V_{vT}(\vec{R}-\beta\vec{r}) + 
H_{vc},
\ee
where
\bea
T_{\vec{R}} &=& \hbar^2 K_R^2 / 2 \mu \qquad \im\vec{K_R}=\del{R} \\*
H_{vc} &=& \hbar^2 K_r^2 / 2 \mu_{vc} + V_{vc}(\vec{r}) \qquad 
\im\vec{K_r}=\del{r}.
\eea
Here, $H_{vc}$ is the internal Hamiltonian of the projectile which has a ground 
state wavefunction, $\phi_0$, satisfying
\be
(\hvcpe) \phi_0 = 0,
\ee
where $\varepsilon_0$ is the binding energy of the projectile.
The core-target and valence-target potentials are $V_{cT}$ and $V_{vT}$ 
respectively, while $V_{vc}$ is the core-valence potential that binds the 
projectile.
The reduced masses of the projectile-target and core-valence systems are $\mu$ 
and $\mu_{vc}$ respectively, while the mass ratios $\alpha=m_v/m_P$ and 
$\beta=-m_c/m_P=\alpha-1$ relate to the distance of the core and valence 
clusters from the center of mass of the projectile.

The exact wavefunction for the few-body scattering system is \cite{glockle}
\bea \label{wf-exact}
\ket{\Psip}&=&\frac{\im\epsilon}{E^+-H}\ket{\phi_0,\vec{K}}\\*
E^+&=&E+\im\epsilon \qquad \epsilon\to0^+,
\eea
where $E$ is the total energy of the few-body scattering system and $\vec{K}$ is 
the centre of mass momentum.
The adiabatic approximation \cite{adia} assumes that the internal motion of the 
projectile is frozen, and thus the internal Hamiltonian of the projectile can be 
replaced by a constant.
This is chosen to be the ground state energy of the projectile 
($-\varepsilon_0$), because then the incident part of the three-body 
wavefunction has unit amplitude.
The adiabatic wavefunction is therefore
\be
\ket{\Psiadp}=\frac{\im\epsilon}{E_0^+-T_{\vec{R}}-V_{cT}-V_{vT}} 
\ket{\phi_0,\vec{K}},
\ee
where $E_0=E+\varepsilon_0$ is the centre of mass energy of the projectile.

\subsection{Core recoil model}
A special case of the adiabatic model is when the scattering is dominated by the 
core-target interaction.
This is the core recoil model in which the valence-target interaction is 
neglected.
This provides a considerable simplification of the adiabatic Hamiltonian, as the 
only place that the vector $\vec{r}$ now appears is in the core-target 
potential.
This dependence can be transformed away using the translation operator
\be
U_R(\vec{x}) = \exp{-\im\vec{x}\cdot\vec{K_R}}.
\ee
The wavefunction in the core recoil model is \cite{crm-prl}
\be
\Psiadp(\Rr) = \phi_0(\vec{r}) \exp{\im\alpha\vec{r}\cdot\vec{K}} \chip(\rcore),
\ee
where $\chip$ is the two-body distorted wave for a particle of mass $\mu$ in the 
core-target potential,
\be
\braket{\vec{R}}{\chip}=\bra{\vec{R}} \im\epsilon \greensad \ket{\vec{K}},
\ee
and $\greensad$ is the adiabatic Green's operator with the valence-target 
potential switched off, and contains the core-target potential in the $\vec{R}$ 
co-ordinate only:
\be
\greensad = \frac{1}{E_0^+-T_{\vec{R}}-V_{cT}(\vec{R})}.
\ee
The core recoil model wavefunction, when evaluated in the appropriate few-body 
$T$-matrix with the valence-target interaction neglected, allows a factorisation 
into a $T$-matrix for a point particle of mass $\mu$, multiplied by a 
formfactor, $F_{00}$ \cite{crm-prl}:
\bea \label{Too-crm}
\tooad\kkp &=& \bra{\vec{K'}} V_{cT} \ket{\chip} F_{00}(\alpha\vec{Q})\\*
F_{00}(\alpha\vec{Q}) &=& \bra{\phi_0} \exp{\im\alpha\vec{r}\cdot\vec{Q}} 
\ket{\phi_0}. \label{Foo}
\eea
The elastic differential cross section can then be obtained from that of a point 
particle multiplied by the formfactor squared:
\be \label{xsec}
\left(\derivof{\Omega}{\sigma}\right)_{\mathrm{el}} = 
\left(\derivof{\Omega}{\sigma}\right)_{\mathrm{point}} 
|F_{00}(\alpha\vec{Q})|^2.
\ee
The point particle elastic cross section is that obtained for a particle of 
reduced mass $\mu$ interacting via the core-target potential, with a centre of 
mass energy $E_0$.
The formfactor includes all effects of excitation and breakup of the halo 
structure.

\subsection{Validity of the adiabatic approximation \label{SEC:VALIDITY}}
The range of validity of the adiabatic approximation has often been investigated 
by comparison with CDCC calculations which do not make the adiabatic 
approximation, but instead, use a discretised continuum spectrum of breakup 
energies for the projectile \cite{deuteron-review}.
By examining the relative excitation and breakup energies required to gain 
convergence in the CDCC calculations it is found that the adiabatic 
approximation gives a better estimate of the elastic cross section than one 
might expect.

Estimates of the average excitation and breakup energies involved in the 
scattering of the projectile have been formulated using the core recoil model 
\cite{nac-greece}.
The adiabatic approximation is expected to be valid when the collision time of 
the projectile and target, $t_{\mathrm{coll}}$, is short in comparison to the 
time associated with the core-valence internal motion, $t_{\mathrm{int}}$.
In Ref.~\cite{nac-greece}, an upper limit on the ratio of these times was given 
as
\be
\frac{t_{\mathrm{coll}}}{t_{\mathrm{int}}} < \lambda,
\ee
where
\be \label{t-ratio}
\lambda = 2\frac{m_v}{m_c}\frac{m_T}{(m_P+m_T)}\frac{R_0}{a^2}\frac{1}{K},
\ee
and $R_0$ defines the range of the core-target interaction and $a$ the 
diffuseness.
The adiabatic approximation is then expected to be valid when
\be \label{lambda}
\lambda \ll 1.
\ee
Note that $\lambda$ is strongly dependent on the valence-to-core mass ratio, and 
only weakly dependent on the incident energy of the projectile, through the 
$1/K$ dependence.
Also note that the derivation of $\lambda$ in Ref.~\cite{nac-greece} involves an 
estimate of the excitation energy involved, but the projectile binding energy 
does not appear.

Using values for the range and diffuseness of the core-target interaction from 
Table~\ref{tab:pot-par}, some typical values for $\lambda$ are
\bea
^{11}\mathrm{Be}+{}^{12}\mathrm{C} \mbox{ at 10 MeV/nucleon}& \Rightarrow& 
\lambda = 0.14,\nonumber\\*
^{6}\mathrm{He}+{}^{12}\mathrm{C} \mbox{ at 10 MeV/nucleon}& \Rightarrow& 
\lambda = 5.\nonumber
\eea
These values set upper limits on the time ratio that must be much less than 
unity.
From these values we would expect that when the valence-target interaction is 
neglected, the adiabatic approximation would give a reasonable description of 
the exact cross section for $^{11}$Be, but a poor description for $^6$He, at 10 
MeV/nucleon.

\begin{table}
\caption{Woods-Saxon potential parameters for core-target and core-valence 
interactions. Energies are in MeV and lengths in fm. The range parameters are to 
be multiplied by the appropriate masses. \label{tab:pot-par}}
\begin{ruledtabular}\begin{tabular}{D{+}{+}{4}|*{6}{r}}
c+T            & $V\;\;\;$    & $R_V\;$   & $a_V\;$   & $W\;\;$     &  $R_W\;$  
& $a_W\;$ \\\hline\rule{0mm}{2.2ex}
{}^{10}\mbox{Be}+{}^{12}\mbox{C} & 123.00 & 0.750 & 0.800 & 65.00 & 0.780 & 
0.800 \\
\alpha+{}^{12}\mbox{C}  &  37.16 & 1.846 & 0.452 & 13.27 & 1.846 & 0.452 
\\\hline\hline
c+v            &  $V\;\;\;$     &  $R_V\;$  &  $a_V\;$  & & & 
\\\hline\rule{0mm}{2.2ex}
{}^{10}\mbox{Be}+n      &  86.42 & 1.000 & 0.530 & && \\
\alpha+2n      & 172.17 & 0.800 & 0.300 & \multicolumn{3}{l}{(ground state)} \\
\alpha+2n      & 134.82 & 0.800 & 0.728 & \multicolumn{3}{l}{($d$-wave 
resonance)}
\end{tabular}\end{ruledtabular}
\end{table}

\subsection{Adiabatic versus CDCC numerical calculations \label{SEC:CDCC}}
Extensive studies of deuteron elastic scattering have shown that there is more 
to the adiabatic approximation than the basic assumption that the excitation 
energy of the projectile is small in comparison to the centre of mass energy of 
the projectile.

\begin{figure}
\includegraphics[width=8.5cm]{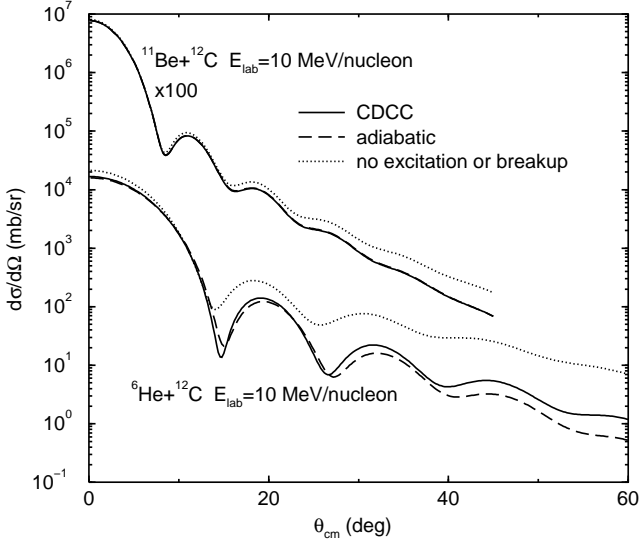}
\caption{CDCC (solid) versus adiabatic (dashed) calculations for $^{11}$Be and 
$^6$He elastic scattering from a $^{12}$C target at 10 MeV/nucleon. The 
valence-target interaction has been neglected as well as the Coulomb 
interaction. The cross section in the limit of no excitation or breakup is 
represented by the dotted line. The $^{11}$Be cross sections are multiplied by 
100. \label{fig:cdcc-adia}}
\end{figure}

Fig.~\ref{fig:cdcc-adia} compares adiabatic and CDCC calculations for $^{11}$Be 
and $^6$He elastic scattering from a $^{12}$C target at 10 MeV/nucleon.
The calculations neglect the valence-target interaction and do not include any 
Coulomb potential.
The adiabatic cross section with the valence-target interaction neglected is 
just the core recoil model cross section obtained from Eq.~(\ref{xsec}).

The core-target potential parameters are given in Table~\ref{tab:pot-par}.
The $^{10}$Be radius parameters are to be multiplied by $10^{1/3}+12^{1/3}$ and 
the $\alpha$ by $12^{1/3}$.

The ground state wavefunction of $^{11}$Be is assumed to be a pure $2s_{1/2}$ 
neutron single particle state, with a separation energy of 0.503 MeV, calculated 
in a central Woods-Saxon potential (Table~\ref{tab:pot-par}).
The potential depth is adjusted to obtain the required binding energy of the 
$^{10}$Be+$n$ system.
Assuming a core root mean squared (rms) radius of 2.28 fm, this generates the 
$^{11}$Be composite nucleus, with rms radius of 2.90 fm, in agreement with a 
recent few-body analysis of halo sizes \cite{halo-radii,halo-radii-l}.

The CDCC calculations were performed using \textsc{fresco} \cite{fresco}.
The $1p_{1/2}$ bound state in $^{11}$Be, with an excitation energy of 320\,keV, 
was included.
The $^{10}$Be+$n$ continuum was modelled using 10 bins from 0-20 MeV for each of 
the $s$-,$p$-,$d$-, and $f$-wave breakup states.

A two-body di-neutron model was assumed for the ground state wavefunction of 
$^6$He, with a binding energy of 0.975 MeV.
The di-neutron is assumed to be in a 2s single particle state. 
The $\alpha$+$2n$ potential used to generate the two-body ground state 
wavefunction was assumed to have a Woods-Saxon form (Table~\ref{tab:pot-par}).

The $\alpha$+$2n$ continuum, due to the breakup of $^6$He, was modelled using 15 
bins from 0-30 MeV for each of the $s$-,$p$-, and $f$-wave breakup states, using 
the same potential as that used for the ground state wavefunction.
The $d$-wave breakup states included a resonance at 1.8 MeV above the ground 
state, with a bin below the resonance and 15 bins up to 30 MeV above the 
resonance.
A potential which reproduced a 1.8 MeV resonance in $\alpha$+$2n$ with a width 
of 113 keV, was used for these $d$-wave breakup states 
(Table~\ref{tab:pot-par}).

The cross section in the limit of no excitation or breakup is calculated by 
folding the core-target potential over the ground state wavefunction, to form a 
two-body projectile-target interaction.
This is shown by the dotted line as a reference to highlight the importance of 
excitation and breakup channels using the different methods.

Fig.~\ref{fig:cdcc-adia} shows that the effect of the adiabatic assumption is 
negligible for $^{11}$Be and small for $^6$He, even though the estimate in 
Section~\ref{SEC:VALIDITY} suggests otherwise.
To understand why this is the case, the leading order corrections to the 
adiabatic approximation are evaluated in the following sections.

\section{First order non-adiabatic corrections \label{SEC:NAC}}
Corrections to the adiabatic approximation arise because the scattering process 
mixes in excited states of the projectile, for which $(\hvcpe)$ is non-zero.
Therefore, we expand the wavefunction of Eq.~(\ref{wf-exact}) in powers of 
$(\hvcpe)$, where to first order we have
\be
\ket{\Psip} = \ket{\Psiadp} + \greensad (\hvcpe) \ket{\Psiadp}.
\ee
The first order non-adiabatic correction to the elastic adiabatic $T$-matrix is 
then
\be \label{deltaT}
\dtooad\kkp = \bra{\Psiadm} (\hvcpe) \ket{\Psiadp}.
\ee

The adiabatic wavefunction is 
\be
\Psiadp(\Rr)=\phi_0(\vec{r})\psiadp(\Rr),
\ee
where $\psiadp$ is the distorted wave for the two-body projectile-target 
scattering system in the potential $V_{cT}(\rcore)+V_{vT}(\rvale)$, with a 
reduced mass $\mu$, and normalised so that it has an incident plane wave in 
$\vec{R}$ with unit amplitude.

The operator product of the 3 factors $\phi_0(\vec{r})$, ($\hvcpe$),\\ and 
$\phi_0(\vec{r})$ which appears in Eq.~(\ref{deltaT}) can be expressed as (see 
Appendix~\ref{APP:OP})
\be \label{id}
\phi_0(\vec{r}) (\hvcpe) \phi_0(\vec{r}) = -\frac{\hbar^2}{2\mu_{vc}} \del{r} 
\phi_0^2 \cdot \del{r}.
\ee
This result assumes that $\phi_0$ is an $s$-state and hence, without loss of 
generality, can be assumed to be real.
The result also assumes that $V_{vc}$ is local.

The expression (\ref{id}) can be conveniently evaluated by allowing the two 
operators to operate in opposite directions, i.e., the left del on the bra and 
the right del on the ket.
Thus, the matrix element is reduced to
\be\begin{split}
\dtooad =& \sum_{\mu=\pm 1,0}(-1)^{\mu} \frac{\hbar^2}{2\mu_{vc}} \\* &\times 
\braket{(\del{r})_{\mu}{\psiadm},\phi_0}{\phi_0,(\del{r})_{-\mu}{\psiadp}}.
\end{split}\ee

\subsection{Non-adiabatic corrections to the core recoil model}
Using the analytical wavefunction of the core recoil model, the matrix element 
for the first order correction to the adiabatic elastic $T$-matrix can be 
simplified significantly.
By translating the two-body distorted waves,
\bea
\chip(\rcore) &=& U_R \chip(\vec{R})\\*
\conj{\chim(\rcore)} &=& \conj{\chim(\vec{R})} U_R^\dagger,
\eea
the adiabatic wavefunctions in the core recoil model can be written as
\bea \label{bra-psi}
\psiadp(\Rr) &=& \eiark \chip(\vec{R})\\*
\conj{\psiadm(\Rr)} &=& \conj{\chim(\vec{R})} \eiarkp. \label{ket-psi}
\eea
The derivatives with respect to $\vec{r}$ are now straight-forward since the 
$\vec{r}$ dependence has been separated.
Therefore, when we take the derivative followed by the inner product of 
Eqs.~(\ref{bra-psi}) and (\ref{ket-psi}), the $\vec{K_R}$ operators in the 
exponents cancel, allowing the formfactor of Eq.~(\ref{Foo}) to be factored out.
This leaves us with the first order correction to the adiabatic elastic 
$T$-matrix:
\be \label{dt-crm}
\dtooad = \frac{\hbar^2\alpha^2}{2\mu_{vc}} \xkkp F_{00}(\alpha\vec{Q})
\ee
where
\be \label{matrix-x}
\xkkp = \bra{\chim} (\vec{K_R}-\vec{K'}) \cdot (\vec{K_R}-\vec{K}) \ket{\chip}.
\ee

\subsection{Evaluation of the non-adiabatic corrections using the eikonal 
approximation}
In Ref.~\cite{nac-jpg}, Eq.~(\ref{dt-crm}) was evaluated using the eikonal 
approximation.
This leads to simple formulae for the corrections.
Using the eikonal approximation to the distorted waves,
\bea
\chip(\vec{R}) &=& \exp{\im\vec{K}\cdot\vec{R}} \exp{-\frac{\im 
K}{2E_0}\int_{-\infty}^{z}\dif z_1 V_{cT}\bzk{z_1}},\\*
\conj{\chim(\vec{R})} &=& \exp{-\im\vec{K'}\cdot\vec{R}} \exp{-\frac{\im 
K'}{2E_0}\int_{z'}^{\infty}\dif z'_1 V_{cT}\bzkp{z'_1}},
\eea
and by making the small angle approximation, $\khat\cdot\kphat=1$, which is 
consistent with the eikonal approximation, Eq.~(\ref{dt-crm}) can be written as 
\cite{nac-jpg}
\be \label{dtoo-eik}
\dtooad = \bra{\chim} \Delta V_{cT} \ket{\chip} F_{00}(\alpha\vec{Q}),
\ee
where
\bwt \be \label{deltavct}
\Delta V_{cT} = \frac{\gamma}{4E_0} \bigg( (V_{cT})^2 -  \int_{-\infty}^{z} \dif 
z_1 \del{b} V_{cT}\bzk{z_1} \cdot \int_{z}^{\infty} \dif z_2 \del{b} 
V_{cT}\bzk{z_2} \bigg),
\ee \ewt
and $\gamma$ is the mass ratio
\be \label{gamma}
\gamma = \frac{m_v}{m_c}\frac{m_T}{(m_P+m_T)}.
\ee

In evaluating the matrix element in Eq.~(\ref{dtoo-eik}), the eikonal distorted 
waves combine to form the eikonal phase factor,
\be \label{chi-ct}
\chi_{cT}(b) = -\frac{K}{2E_0} \int_{-\infty}^{\infty}\dif z V_{cT}\bzk{z},
\ee
in which the $z$ integration has been performed.
Note that this is possible since, for elastic scattering, $K'=K$.
The $z$ integration in Eq.~(\ref{dtoo-eik}) therefore only involves $\Delta 
V_{cT}$, and we can define the quantity
\be \label{chi-tilde-deltaVct}
\tilde{\chi}(b) = - \frac{K}{2E_0} \int_{-\infty}^{\infty}\dif z \Delta V_{cT}.
\ee
For a central potential, the integral over all $z$ of $\Delta V_{cT}$ can be 
simplified to a single integral over all $z$ of the potential squared,
\be \label{chi-tilde}
\tilde{\chi}(b) = - \frac{\gamma K}{8E_0^2} \left(1+b\deriv{b}\right) 
\int_{-\infty}^{\infty} \dif z V_{cT}^2(\sqrt{b^2+z^2}).
\ee
We can now write Eq.~(\ref{dtoo-eik}) as a scattering amplitude,
\be
\Delta f_{00}^{\eik} = K \intb S(b) \tilde\chi(b) F_{00}(\alpha\vec{Q}).
\ee

The eikonal scattering amplitude in the core recoil model, including first order 
non-adiabatic corrections, is then
\be \label{deltafeik}
\bar{f}_{00}^{\eik} = \im K \intb \left[1- S(b)(1+\im\tilde{\chi}) \right] 
F_{00}(\alpha\vec{Q}),
\ee
where $\bar{f}_{00}^{\eik} = f_{00}^{\eik} + \Delta f_{00}^{\eik}$.
Here,
\be
S(b)=\expl{-\frac{\im K}{2E_0} \int_{-\infty}^{\infty}\dif z 
V_{cT}(\sqrt{b^2+z^2})}
\ee
is the eikonal $S$-matrix for $^{11}$Be as a point particle moving in the 
core-target potential.

We see that the non-adiabatic correction term, $\tilde\chi$ of 
Eq.~(\ref{chi-tilde}), is a dimensionless quantity, whose size relative to unity 
determines the magnitude of the corrections.

The structure of the expression (\ref{chi-tilde}) for the correction factor 
$\tilde\chi$ can be traced to the physical origin of corrections to the 
adiabatic approximation.
In the few-body model used here these corrections arise from excitation of the 
projectile through tidal forces generated by the  interaction of the core with 
the target.
These forces arise because of the displacement of the core from the 
centre-of-mass of the projectile and it is therefore natural 
that the mass ratio $m_v/m_c$ and derivatives of the core-target potential 
should be crucial.
There is a quadratic dependence on $V_{cT}$ because a two-step process must be 
involved if the projectile is to end up in the ground state (elastic 
scattering).
The extra $1/E_0$ factor in $\tilde\chi$, over and above that expected from the 
energy dependence of the eikonal phase (see Eqs.~(\ref{chi-ct}) and 
(\ref{chi-tilde-deltaVct})), must be related to the expected dependence on the 
collision time discussed in Section~\ref{SEC:VALIDITY}.

We note that $\tilde\chi$ is multiplied by the eikonal $S$-matrix which will 
restrict the impact parameters contributing to the overall corrections.

\subsection{Exact evaluation of non-adiabatic corrections \label{SEC:EXACT}}
While the eikonal approximation provides useful insights into the nature of the 
non-adiabatic corrections, the matrix element in Eq.~(\ref{matrix-x}) can be 
evaluated exactly using a partial wave expansion.
The quantity $\xkkp$, defined in Eq.~(\ref{matrix-x}), was shown in 
Ref.~\cite{nac-jpg} to have the alternative form,
\be \label{matrix-alt}
\xkkp = -\bra{\chim} \comkv \greensad{}^2 \comkv \ket{\chip}.
\ee

We can write $\chip$ as a partial wave sum,
\be \label{chi-pw}
\braket{\vec{R}}{\chip} = 4\pi \sum_{\ell m} \im^\ell \spharmc{\ell m}{K} 
\spharm{\ell m}{R} \chil(R),
\ee
where $\chil$ is the solution of the radial Schr\"{o}dinger equation
\be \label{homo}
(E_0-T_{\ell}-V_{cT}) \chil(R) = 0,
\ee
and $T_\ell$ is radial kinetic energy operator,
\be
T_\ell = - \cona \frac{1}{R} \dderiv{R} R + \cona \frac{\ell(\ell+1)}{R^2}.
\ee
The partial distorted wave, $\chil$, has the asymptotic form (ignoring the 
Coulomb interaction)
\be \label{chil-asym}
\chil(R) \asym \frac{\im}{2KR} \left( H_{\ell}^{(-)} - S_\ell H_{\ell}^{(+)} 
\right),
\ee
where $S_\ell$ is the partial wave $S$-matrix, and $H_{\ell}^{(\pm)}$ are radial 
Hankel functions, i.e., irregular solutions of the equation
\be
\left[ \dderiv{R} - \frac{\ell(\ell+1)}{R^2} - \frac{2\mu}{\hbar^2} V + K^2 
\right] y_\ell = 0.
\ee 

With a similar expression for $\chim$ and using a partial wave sum for 
$\greensad$, the matrix element of Eq.~(\ref{matrix-alt}) can be reduced to 
\cite{phd-ncs}
\be
\xkkp = 4\pi \sum_\ell P_\ell(\cos\theta) \xexact,
\ee
where we have defined
\be\begin{split} \label{xl}
\xexact = \Bigg[(\ell+1) & \intr{0}{\infty} 
\left(\chibara{K,\ell,\ell+1}\right)^2 \\* {}+ \ell & \intr{0}{\infty} 
\left(\chibara{K,\ell,\ell-1}\right)^2 \Bigg].
\end{split}\ee
The non-adiabatic correction to the elastic scattering amplitude in the core 
recoil model is then
\be \label{deltaf}
\Delta f_{00} = - \gamma F_{00}(\alpha\vec{Q}) \sum_\ell P_\ell(\cos\theta) 
\xexact.
\ee

In Eq.~(\ref{xl}), $\chibarl$ is the solution of the inhomogeneous equation
\be \label{inhom}
(E_0-T_{\ell'}-V_{cT}) \chibarl(R) = \derivof{R}{V_{cT}} \chil(R),
\ee
with asymptotic form (ignoring the Coulomb interaction)
\be \label{chibar-asym}
\chibarl(R) \asym \frac{\im}{2R} S_{\ell\ell'} H_{\ell'}^{(+)},
\ee
and where $\ell'$ can take two values, which leads to the two terms in 
Eq.~(\ref{xl}):
\be
\ell' = \ell \pm 1.
\ee
We have defined $S_{\ell\ell'}$ as the coefficient of the Hankel function in the 
inhomogeneous solution.
It can be shown that it is equivalent to the difference between the $S$-matrix 
for the homogeneous solutions for $\ell$ and $\ell'$,
\be
S_{\ell\ell'} = \pm(S_{\ell} - S_{\ell'}) \quad:\quad \ell'=\ell\pm1.
\ee

The oscillatory nature of $\chibarl(R)$, for large $R$, means that the integrals 
in Eq.~(\ref{xl}) have to be dealt with carefully.
The asymptotic form for the solutions to the differential Eqs.~(\ref{homo}) and 
(\ref{inhom}) is reached when $R$ is outside the range of the potential.
The radial Hankel functions, $H_\ell^{(\pm)}$, have the asymptotic form
\be
H_\ell^{(\pm)} \asym \exp{\pm\im(KR-\ell\pi/2)},
\ee
when
\be \label{asym-cond}
KR\gg\ell(\ell+1).
\ee
The solutions of the differential equations were computed out to a radius 
$R=R_0$, which is outside the range of the potential.
Then, the integrals from $R_0$ to $R_\ell$, where $R_\ell$ is chosen so that the 
asymptotic condition (\ref{asym-cond}) is met, are performed using the explicit 
form of the Hankel function \cite{messiah1},
\be
H_\ell^{(+)} = \exp{\im KR} \sum_{s=0}^\ell \frac{\im^{s-\ell}}{2^ss!} 
\frac{(\ell+s)!}{(\ell-s)!} (KR)^{-s}.
\ee
The integral of the asymptotic form of the Hankel function from 
$R_\ell\to\infty$ can be performed analytically by making the substitution 
$R=\im y +R_\ell$, and integrating over the complex plane,
\bea
\int_{R_{\ell'}}^{\infty}\dif R \exp{2\im(KR-\ell'\pi/2)} &=& \im 
\int_{0}^{\infty}\dif y \exp{-2Ky+2\im KR_{\ell'}-\im\ell'\pi} \nonumber\\* &=& 
\frac{\im}{2K}\exp{2\im KR_{\ell'}-\im\ell'\pi}.
\eea

It it well known that a partial wave sum can be written exactly as an integral 
over impact parameters \cite{jak-jat-scat}.
By making the semi-classical correspondence, $\ell=bK$, and assuming the 
scattering angle is small, the sum over partial waves can be written as 
\cite{sakurai}
\be
\sum_\ell P_\ell(\cos\theta) \to K\int_0^\infty \dif b J_0(Qb).
\ee
The eikonal equivalent of $\xexact$ (Eq.~(\ref{xl})) can then be found by 
substituting this relation into Eq.~(\ref{deltaf}) and comparing it with 
Eq.~(\ref{deltafeik}) to obtain
\be \label{Xb}
\gamma\xexact \equiv b S(b) \tilde{\chi}(b),
\ee
where $\gamma$ is the mass ratio of Eq.~(\ref{gamma}), which also appears on the 
R.H.S. within the correction term, $\tilde\chi$.
It is then understood that it is the magnitude of either side of Eq.~(\ref{Xb}) 
which determines the overall size of the non-adiabatic corrections: it contains 
the overlap of the correction term with the elastic scattering $S$-matrix, which 
determines which impact parameters contribute to the cross section.

\section{Numerical results \label{SEC:RESULTS}}
\subsection{Application to $^{11}$Be+$^{12}$C elastic scattering}
Here we examine the non-adiabatic corrections in the core recoil model for 
$^{11}$Be scattering from a $^{12}$C target.
$^{11}$Be is a good example of a single neutron halo nucleus to which the core 
recoil model can be applied.
The interaction between the $^{10}$Be core and the target dominates the elastic 
scattering and it is thus a reasonable first approximation to neglect the 
neutron-target interaction.
The core recoil model was applied to the elastic scattering of 
$^{11}$Be+$^{12}$C at 49.3 MeV/nucleon in Ref.~\cite{crm-prl}.
At this energy we expect the adiabatic approximation to be good, and we 
therefore calculate the non-adiabatic corrections at 10 MeV/nucleon, where we 
have calculated the qualitative estimates for the validity of the approximation 
earlier.

The potentials and wavefunctions were discussed in Section~\ref{SEC:CDCC}. 
The potential used for the $^{10}$Be core in Ref.~\cite{crm-prl} was obtained 
from elastic scattering data at 59.4 MeV/nucleon.
Due to the absence of data at lower energies, the same potential will be used at 
10 MeV/nucleon, but as the corrections are dependent on the potential geometry, 
ideally the potential should be fixed by elastic scattering of $^{10}$Be at 10 
MeV/nucleon.

\begin{figure}
\includegraphics[width=8.5cm]{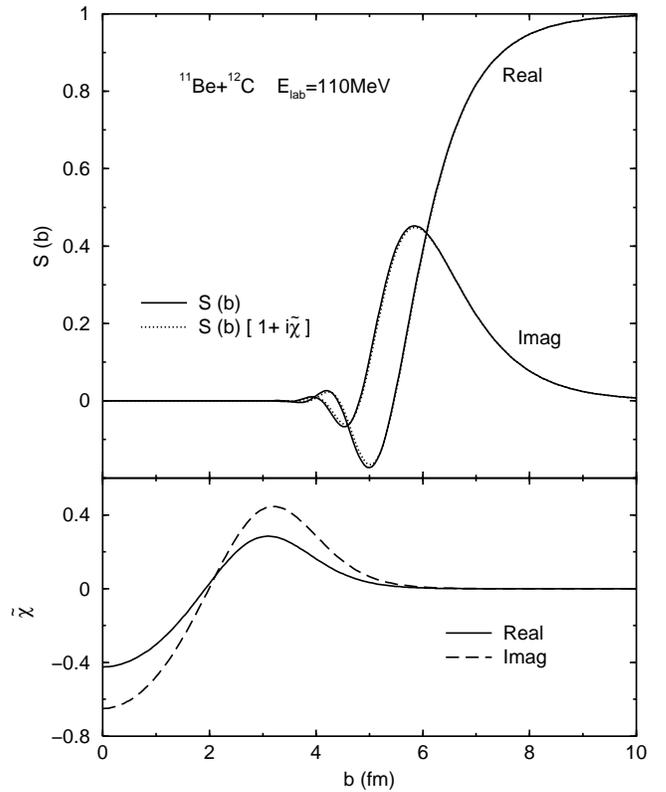}
\caption{Eikonal $S$-matrix and non-adiabatic corrections for $^{11}$Be+$^{12}$C 
in the core recoil model. In the top figure, the solid line represents the 
two-body $S$-matrix for $^{11}$Be as a point particle interacting with the 
$^{12}$C target via the core-target interaction. The dotted line includes the 
first order non-adiabatic corrections through multiplying the $S$-matrix by the 
factor $(1+\im\tilde{\chi})$, while the correction term, $\tilde{\chi}$, is 
plotted in the bottom figure on the same impact parameter scale. 
\label{fig:be11-10.sbar}}
\end{figure}

\begin{figure}
\includegraphics[width=8.5cm]{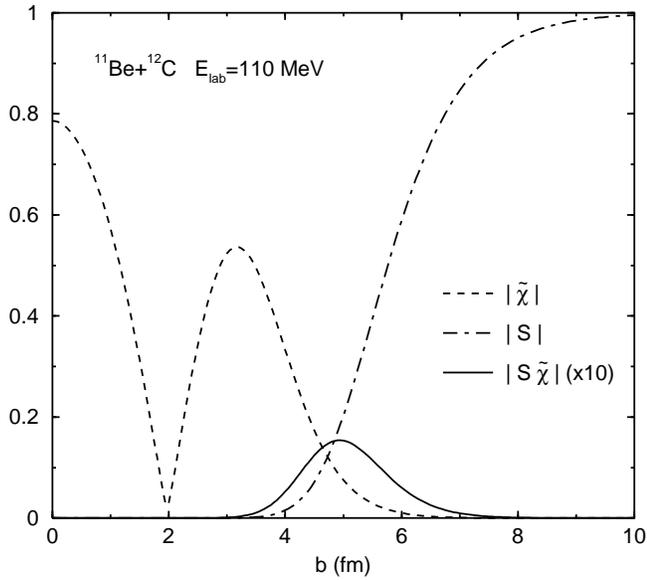}
\caption{Overlap of $S$-matrix and non-adiabatic corrections for 
$^{11}$Be+$^{12}$C. The dashed line is the magnitude of the correction term 
$\tilde\chi$ plotted against impact parameter. The dot-dashed line is the 
magnitude of the $S$-matrix. The solid line is the magnitude of the product of 
the $S$-matrix and the correction term $\tilde\chi$ multiplied by 10. 
\label{fig:be11.overlap}}
\end{figure}

In Fig.~\ref{fig:be11-10.sbar} (bottom), the correction term, $\tilde\chi$, is 
plotted against impact parameter for $^{11}$Be+$^{12}$C at 10 MeV/nucleon.
It has a maximum value of approximately 0.6, which is a significant correction 
as it is added to unity, but it is then multiplied by the eikonal $S$-matrix, 
which is plotted on the same impact parameter scale in 
Fig.~\ref{fig:be11-10.sbar} (top).
We see that the non-adiabatic corrections are largest for small impact 
parameters, but the $S$-matrix is zero in this region, due to a large imaginary 
component in the core-target potential.
This kills off most of the correction term, with only large impact parameters, 
corresponding to grazing collisions, contributing to the cross section.
The large impact parameters correspond to forward scattering angles where there 
is little momentum transferred to the projectile during the scattering, and 
therefore only small corrections to the adiabatic approximation.
This can be seen more clearly in Fig.~\ref{fig:be11.overlap}, where the 
magnitude of the $S$-matrix and correction term, $\tilde\chi$, is plotted along 
with the overlap of the two functions (multiplied by 10).
It is the maximum size of this overlap, $|S\tilde\chi|_{\mathrm{max}}$, in 
comparison to unity, which determines the overall size of the non-adiabatic 
corrections.
We see that for this system, the maximum value is 0.015, so the non-adiabatic 
corrections are small.

\subsection{Accuracy of eikonal calculations}
The eikonal approximation provides us with an understanding to the nature of the 
non-adiabatic corrections, but the use of the eikonal approximation must be 
validated, as in the energy region we are considering, we would expect 
significant non-eikonal corrections.

The non-adiabatic corrections were formulated exactly in 
Section~\ref{SEC:EXACT}; by comparison of the exact non-adiabatic corrections 
with those calculated in the eikonal approximation, the validity of the eikonal 
approximation for these calculations can be assessed.

Eq.~(\ref{Xb}) shows that $\gamma\xexact$ can be compared to the overlap of the 
correction term and the $S$-matrix in the eikonal approximation, as shown in 
Fig.~\ref{fig:be11.overlap}, multiplied by $b$.

\begin{figure}
\includegraphics[width=8.5cm]{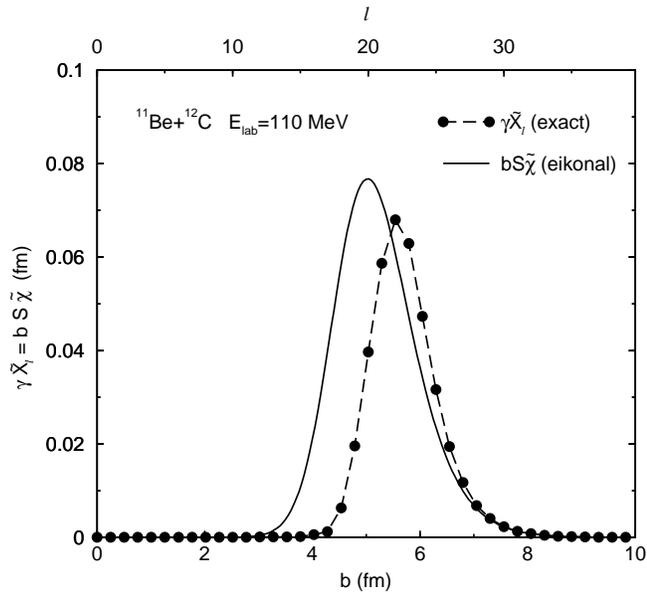}
\caption{Exact and eikonal calculations of non-adiabatic corrections for 
$^{11}$Be+$^{12}$C at 10 MeV/nucleon. The solid line represents the eikonal 
approximation to the non-adiabatic corrections plotted against impact parameter 
along the bottom axis. It is the overlap of the correction term and the 
$S$-matrix weighted by the corresponding impact parameter. The circles represent 
the exact non-adiabatic corrections for each partial wave, which are labelled 
along the top axis and scaled to match the corresponding impact parameter via 
the relation $\ell=bK$, where $K=3.97$ fm$^{-1}$. The exact correction term, 
$\xexact$, is scaled by the mass ratio $\gamma$. \label{fig:be11.eve}}
\end{figure}

Each side of Eq.~(\ref{Xb}) is plotted in Fig.~\ref{fig:be11.eve}, with the 
exact calculation (L.H.S. of Eq.~(\ref{Xb})) represented by the circles, and the 
eikonal calculation (R.H.S. of Eq.~(\ref{Xb})) represented by the line.
The exact calculation is plotted for each partial wave and scaled to match the 
corresponding impact parameter for the eikonal calculations.
This comparison shows that the overlap of the eikonal $S$-matrix and the 
correction term, $\tilde\chi$, gives a reasonable representation of the 
non-adiabatic corrections.
The corrections are slightly over-estimated by the eikonal approximation, 
especially for the smaller impact parameters, but the larger impact parameters 
are well repreduced.
This is where we expect the eikonal approximation to do better as this 
corresponds to smaller scattering angles.

\begin{figure}
\includegraphics[width=8.5cm]{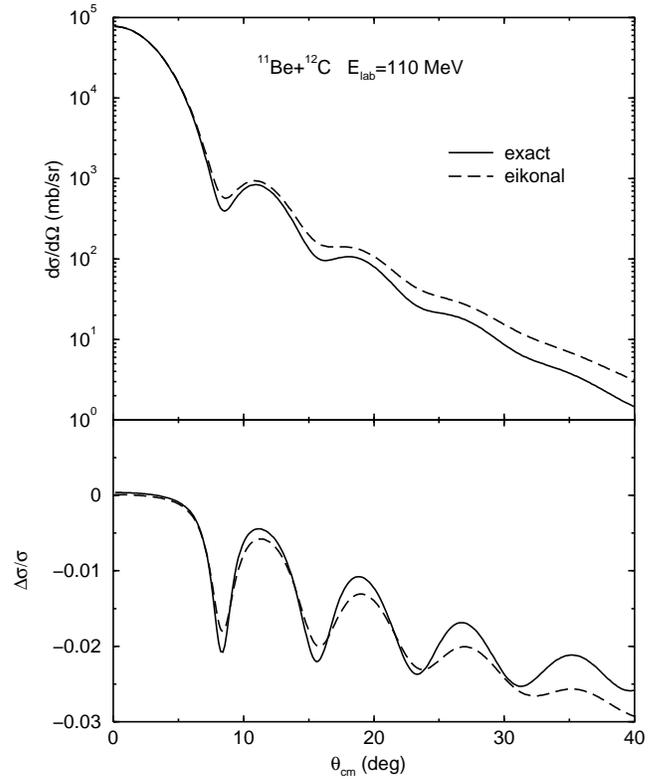}
\caption{Angular distribution of the elastic differential cross section for 
$^{11}$Be+$^{12}$C in the core recoil model with no Coulomb interaction. The 
upper figure compares the cross sections in the core recoil model, using exact 
(solid) and eikonal (dashed) calculations of the two-body cross section. The 
lower figure shows the non-adiabatic correction as a fraction to the cross 
section, with the lines corresponding to exact (solid) and eikonal (dashed) 
calculations. \label{fig:be11-10.xsec}}
\end{figure}

In Fig.~\ref{fig:be11-10.xsec} (top), the differential cross section in the core 
recoil model is plotted.
The two curves represent the different methods of calculating the two-body cross 
section in the core recoil model.
The solid line uses an exact partial wave analysis while the dashed line uses 
the eikonal approximation; the formfactor is the same in both cases.
In the bottom of Fig.~\ref{fig:be11-10.xsec} the first order non-adiabatic 
correction is plotted as a fraction to the core recoil cross section.
We see that the fractional correction to the adiabatic approximation is well 
reproduced by the eikonal approximation, even though the eikonal cross section 
differs significantly from the exact cross section in the core recoil model.
The eikonal approximation over-estimates the cross section at this energy, but 
we saw from Fig.~\ref{fig:be11.eve} that it also over-estimates the magnitude of 
the corrections.
As a fraction to the cross section, however, the non-adiabatic corrections are 
well reproduced.
The large core absorption, which kills of most of the corrections, means that 
only large impact parameters contribute to the cross section, and therefore, the 
eikonal approximation gives a reasonable description of the non-adiabatic 
corrections.

We note that the fractional corrections to the adiabatic approximation do not 
depend on the formfactor, and therefore do not depend on the internal structure 
of the projectile.
This is a result of our assumption of an $s$-wave projectile which results in 
the factorisation shown in Eq.~(\ref{dt-crm}).
For a non-$s$-wave projectile, the additional terms in Eq.~(\ref{id1}) 
contribute.

In Section~\ref{SEC:VALIDITY}, an estimate on the accuracy of the adiabatic 
approximation was given as an upper limit on a time ratio, which had to be much 
less than one.
This upper limit was calculated to be 0.14 for $^{11}$Be+$^{12}$C at 10 
MeV/nucleon: we then would expect the adiabatic approximation to be good at this 
energy from this estimate.
From the calculations of the first order corrections, the adiabatic 
approximation is extremely accurate for this system, going beyond the range that 
the estimate of Section~\ref{SEC:VALIDITY}, due to the key role of the strong 
absorption associated with the scattering at small impact parameters.
The maximum overlap function had a value of 0.015 at the peak, which suggests 
that the adiabatic approximation is approximately 10 times better that the 
qualitative estimate of Section~\ref{SEC:VALIDITY}.

\subsection{Application to $^6$He+$^{12}$C elastic scattering}
In the previous section, the $^{11}$Be+$^{12}$C system was studied because it 
was a reasonable approximation to use the core recoil model, as the ratio of the 
valence-to-core masses was $1/10$.
This small ratio also meant that the corrections in this model were small.
The large absorption in the $^{10}$Be+$^{12}$C potential also played an 
important role in accuracy of the adiabatic approximation for elastic 
scattering.
To examine the role of the mass ratio and core absorption, the core recoil model 
was applied to $^6$He+$^{12}$C elastic scattering.

The $^6$He nucleus has a two-neutron halo with an $\alpha$ core, so the ratio of 
valence-to-core masses is $1/2$.
The $\alpha$ core is light and will appear slightly transparent to the $^{12}$C 
target, so corrections at small impact parameters will contribute.
Elastic scattering cross sections are available over a wide range of energies 
for $\alpha$+$^{12}$C, so a more realistic potential can be used.
The potentials and wavefunctions were discussed in Section~\ref{SEC:CDCC}. 

\begin{figure}
\includegraphics[width=8.5cm]{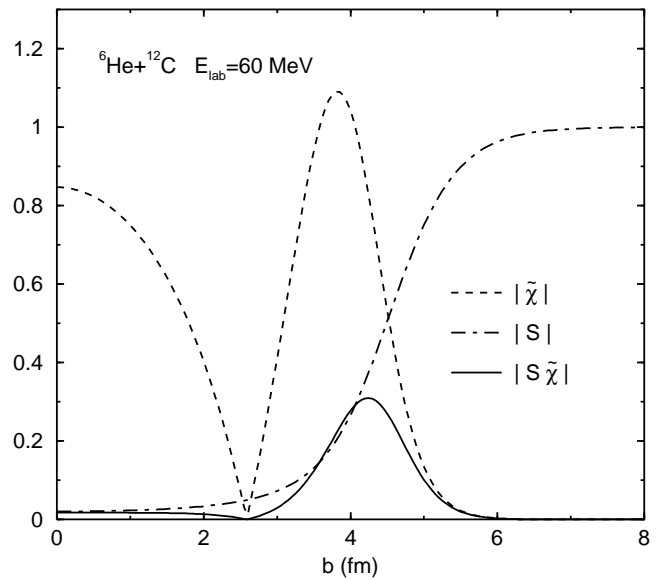}
\caption{Overlap of $S$-matrix and non-adiabatic corrections for 
$^6$He+$^{12}$C. The dashed line is the magnitude of the correction term 
$\tilde\chi$ plotted against impact parameter. The dot-dashed line is the 
magnitude of the $S$-matrix. The solid line is the magnitude of the product of 
the $S$-matrix and the correction term $\tilde\chi$. \label{fig:he6.overlap}}
\end{figure}

The magnitude of the non-adiabatic correction term, $\tilde\chi$, is plotted in 
Fig.~\ref{fig:he6.overlap} (dashed line).
We see that the $b=0$ value is approximately the same as for $^{11}$Be at the 
same energy.
Despite the valence-to-core mass ratio being five times larger for $^6$He, the 
smaller $\alpha$+$^{12}$C potential cancels this out, leaving a correction term 
similar to that for $^{11}$Be.
The magnitude of the peak at the potential surface for $\tilde\chi$ is larger 
than its $b=0$ value; for $^{11}$Be it was smaller.
This is because the $\alpha$+$^{12}$C potential has a sharper potential surface, 
increasing the derivative of the potential at the surface, thus increasing the 
correction term (see Eq.~(\ref{chi-tilde})).
By comparison with the $S$-matrix (dot-dashed line), we see that the peak in the 
correction term now appears in a region of the $S$-matrix which is non-zero; 
therefore, the overlap with the $S$-matrix is more significant.
It has a maximum overlap of around 0.3, compared to 0.015 in the $^{11}$Be case.
This factor of 20 difference between the two cases arises even though the 
magnitude of $\tilde\chi$ at $b=0$ is of the same order in both cases.

The overlap between the correction term, $\tilde\chi$, and the $S$-matrix is 
increased in the $^6$He case because the core-target potential has a weaker 
absorption associated with it.
This increases the magnitude of the $S$-matrix in the region of maximum 
corrections, thus increasing the overlap.
Even so, the $S$-matrix is still relative small at the peak of the correction 
term, $\tilde\chi$, and so the maximum overlap is still much smaller than the 
maximum size of the correction term.
The maximum overlap of 0.3 is small in comparison to unity, thus corrections to 
the adiabatic approximation are still small, even though the qualitative 
estimate in Section~\ref{SEC:VALIDITY} suggested that the approximation would be 
poor at this energy.

\begin{figure}
\includegraphics[width=8.5cm]{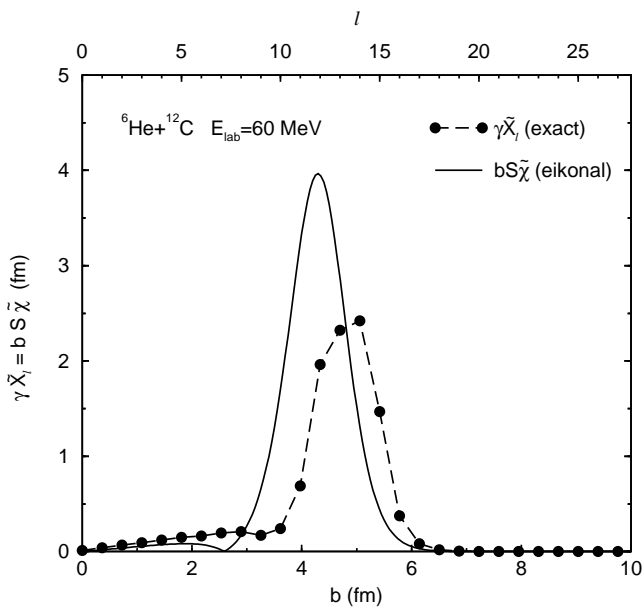}
\caption{Exact and eikonal calculations of non-adiabatic corrections for 
$^{11}$Be+$^{12}$C at 10 MeV/nucleon. The solid line represents the eikonal 
approximation to the non-adiabatic corrections plotted against impact parameter 
along the bottom axis. It is the overlap of the correction term and the 
$S$-matrix weighted by the corresponding impact parameter. The circles represent 
the exact non-adiabatic corrections for each partial wave, which are labelled 
along the top axis and scaled to match the corresponding impact parameter via 
the relation $\ell=bK$, where $K=2.77$ fm$^{-1}$. The exact correction term, 
$\xexact$, is scaled by the mass ratio $\gamma$. \label{fig:he6.eve}}
\end{figure}

\begin{figure}
\includegraphics[width=8.5cm]{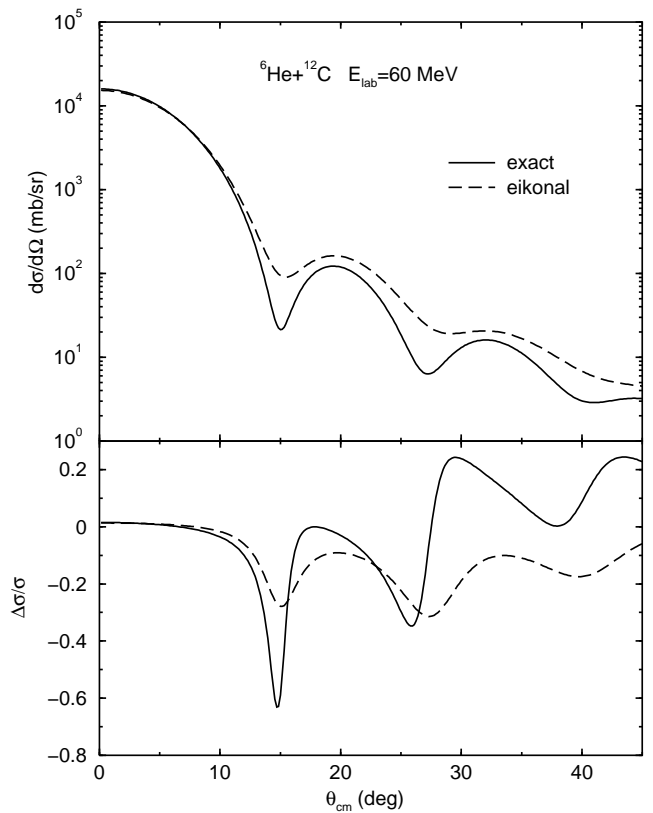}
\caption{Angular distribution of the elastic differential cross section for 
$^6$He+$^{12}$C in the core recoil model with no Coulomb interaction. The upper 
figure compares the cross sections in the core recoil model, using exact (solid) 
and eikonal (dashed) calculations of the two-body cross section. The lower 
figure shows the non-adiabatic correction as a fraction to the cross section, 
with the lines corresponding to exact (solid) and eikonal (dashed) calculations. 
\label{fig:he6-10.xsec}}
\end{figure}

The accuracy of eikonal calculations for the $^6$He case is shown in 
Fig.~\ref{fig:he6.eve}.
We see, as in the $^{11}$Be case, that the non-adiabatic corrections are 
over-estimated by the eikonal approximation, with the worst agreement for 
smaller impact parameters, but we see in Fig.~\ref{fig:he6-10.xsec} that this 
difference is more evident in the fractional correction to the cross section.
The $S$-matrix is not as strongly absorbing as for $^{11}$Be, so these small 
impact parameters are contributing to the cross section at large scattering 
angles.
For the largest scattering angles (above 25$^\circ$), the eikonal approximation 
to the first order non-adiabatic corrections reduce the cross section, while an 
exact evaluation shows an increase in the cross section.
The eikonal approximation was good for $^{11}$Be because the large core 
absorption meant that only large impact parameters contributed.
This is not the case for $^6$He scattering and so the eikonal approximation is 
not as accurate.

\subsection{Dependence on core absorption}
In the previous section, we see that the non-adiabatic corrections 
$^6$He+$^{12}$C at 10 MeV/nucleon are larger than for $^{11}$Be+$^{12}$C at the 
same energy.
The reasons for this are two fold: firstly, the core-to-valence mass ratio is 
much larger for the former; secondly, the core-target potential has a smaller 
imaginary component for the former, due to less absorption of the core.
The second effect is examined in more detail here.

The potential we have used for the $^{10}$Be+$^{12}$C interaction was obtained 
at 59.4 MeV/nucleon.
As the non-adiabatic corrections have been calculated at the energy of 10 
MeV/nucleon, we would expect that the imaginary potential strength to be 
reduced, but without any experimental data at this energy, its precise value 
cannot be fixed.

\begin{figure}
\includegraphics[width=8.5cm]{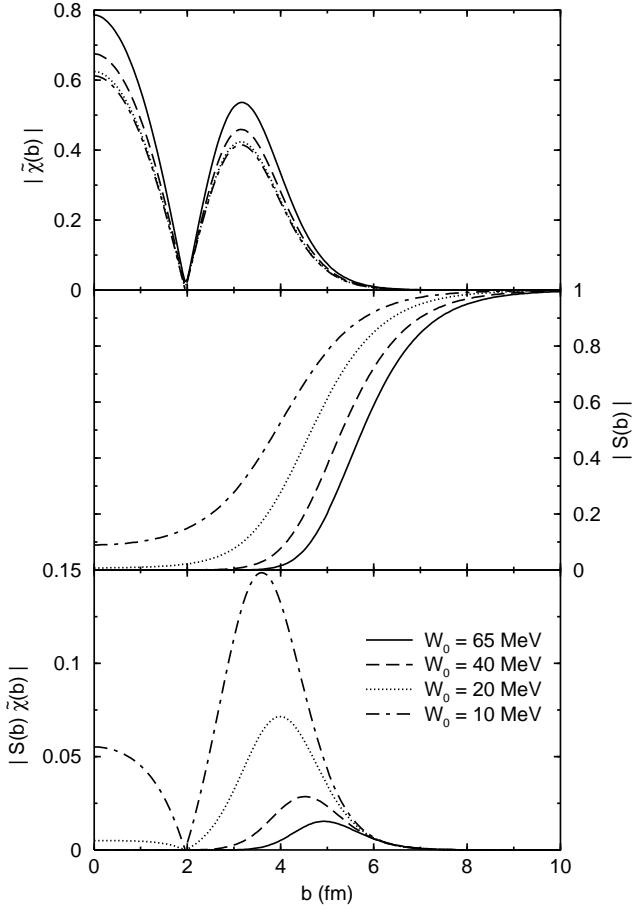}
\caption{Non-adiabatic correction term for $^{11}$Be+$^{12}$C at 10 MeV/nucleon 
for varying imaginary potential depths. The top figure plots the modulus of the 
correction term, $\tilde\chi$, versus impact parameter while the middle figure 
shows the modulus of the $S$-matrix on the same impact parameter scale. The 
bottom figure is the overlap of the $S$-matrix and $\tilde\chi$. The imaginary 
potential depths for the core-target potential are given in the legend. 
\label{fig:overlap}}
\end{figure}

The dependence of the corrections on the imaginary potential strength, $W_0$, is 
shown in Fig.~\ref{fig:overlap}.
The non-adiabatic corrections are plotted for a range of imaginary potential 
strengths for the $^{11}$Be+$^{12}$C reaction at 10 MeV/nucleon.
Fig.~\ref{fig:overlap} contains three graphs: the top graph plots the modulus of 
the correction term, $\tilde\chi$; the middle figure is the modulus of the 
$S$-matrix; the bottom figure is the overlap of the two.
The dependence of the corrections on the imaginary potential is shown for 
various imaginary potential strengths: the solid line represents the 65 MeV 
imaginary potential that was obtained from the elastic scattering of 
$^{10}$Be+$^{12}$C at 59.4 MeV/nucleon, the dashed line corresponds to a 40 MeV 
imaginary potential, the dotted line is for $W_0$=20 MeV, and the dot-dashed 
line represents $W_0$=10 MeV.

We see that different imaginary potential strengths do not effect the correction 
term $\tilde\chi$ significantly, but the effect on the $S$-matrix is large.
As the imaginary potential strength is reduced, the correction term, 
$\tilde\chi$, is reduced slightly; but, the target appears more transparent and 
so the $S$-matrix is increased significantly for the small impact parameters.
The overlap of the correction term with the $S$-matrix is then significantly 
increased, as shown in the bottom figure, and it is this overlap which 
determines the overall size of the corrections.

\begin{figure}
\includegraphics[width=8.5cm]{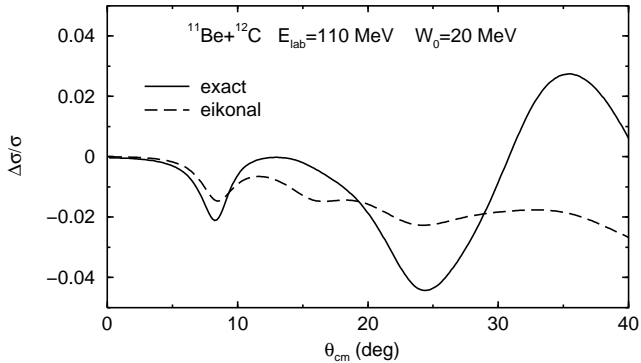}
\caption{Angular distribution of the elastic differential cross section for 
$^{11}$Be+$^{12}$C in the core recoil model with the imaginary potential depth 
reduced to 20 MeV. The non-adiabatic corrections are plotted as a fraction to 
the cross section, with the lines corresponding to exact (solid) and eikonal 
(dashed) calculations. \label{fig:be11-w20.xsec}}
\end{figure}

The accuracy of the eikonal approximation for evaluating the non-adiabatic 
corrections was shown to be suspect for $^6$He, which was said to be due to the 
weak imaginary potential.
To see this effect for the $^{11}$Be case, the 20 MeV imaginary potential was 
used to compare the exact and eikonal calculations of the non-adiabatic 
corrections.
The fractional non-adiabatic corrections are shown in 
Fig.~\ref{fig:be11-w20.xsec}.
We see here that the eikonal approximation reproduces the exact corrections 
poorly for large scattering angles (above 20$^\circ$), as in the $^6$He case.
The eikonal approximation fails to predict the increase in the cross section for 
large scattering angles that was seen when the exact non-adiabatic corrections 
were included.
The non-adiabatic corrections for $^{11}$Be are still small compared to the 
$^6$He calculations, even when the imaginary potential depth is reduced, due to 
the smaller valence-to-core mass ratio.

\section{Improved estimate for validity of adiabatic approximation 
\label{SEC:NEW-EST}}
We have shown in our numerical evaluation of the non-adiabatic corrections that 
strong core absorption improves the accuracy of the adiabatic approximation, 
going beyond that expected from the estimates of Section~\ref{SEC:VALIDITY}.

The key point is that when strong absorption is present, it is the value of the 
correction term, $\tilde\chi$ (Eq.~(\ref{chi-tilde})), in the region near the 
strong absorption radius which is of importance.
The size of the non-adiabatic corrections is determined by the maximum overlap 
of $\tilde\chi$ and the $S$-matrix.
If you model $\tilde\chi$ by an exponential and the $S$-matrix by a Woods-Saxon, 
then a good estimate for the size of the non-adiabatic corrections is 
$\tilde\chi$, evaluated at the strong absorption radius, multiplied by one half 
(the value of the $S$-matrix at the strong absorption radius).

We can estimate the value of $\tilde\chi$ (Eq.~(\ref{chi-tilde})) at the strong 
absorption radius as the potential in this region has the form of a simple 
exponential, so that
\be
\int_{-\infty}^{\infty}\dif z V_{cT}^2(\sqrt{b^2+z^2}) = 
\int_{-\infty}^{\infty}\dif z V_{0}^2 \exp{-2\sqrt{b^2+z^2}/a},
\ee
where $a$ is the diffuseness of the core-target interaction.
Since the most important values in the $z$ integral are those around $z=0$, the 
square root can be expanded in powers of $z/b$:
\be
\int_{-\infty}^{\infty}\dif z V_{0}^2 \exp{-2\sqrt{b^2+z^2}/a} \approx 
\int_{-\infty}^{\infty}\dif z V_{0}^2 \exp{-2b(1+z^2/2b^2)/a}.
\ee
The integral of the square of the potential can therefore be written
\be
\int_{-\infty}^{\infty}\dif z V_{cT}^2(\sqrt{b^2+z^2}) \approx \sqrt{ba\pi} 
V_{cT}^2(b),
\ee
and an estimate of $\tilde\chi$, near the strong absorption radius, is
\be
\tilde\chi(b) \approx \frac{\gamma K}{8E_0^2} \sqrt{a\pi} \deriv{b} 
b^{3/2}V_{cT}^2(b).
\ee
The derivative of the potential dominates in $\deriv{b} b^{3/2}V_{cT}^2(b)$, and 
as the potential only depends on $b$, the overlap of the correction term and the 
$S$-matrix at the strong absorption radius can be written,
\be \label{new-est}
|S\tilde\chi|_{\mathrm{max}} \approx \frac{\gamma K}{16E_0^2} \sqrt{\pi a} 
R_{\mathrm{s}}^{3/2} \left.\deriv{R} V_{cT}^2(R)\right|_{R=R_{\mathrm{s}}},
\ee
where $R_{\mathrm{s}}$ is the strong absorption radius, and $\gamma$ is the mass 
ratio in Eq.~(\ref{gamma}).
The adiabatic approximation, for scattering systems with strong core absorption, 
is then valid when
\be
|S\tilde\chi|_{\mathrm{max}} \ll 1.
\ee
For strong absorbing systems, this criterion replaces condition~(\ref{lambda}).

Eq.~(\ref{new-est}) gives approximate values for $|S\tilde\chi|_{\mathrm{max}}$ 
of 0.01 and 0.2 for $^{11}$Be and $^6$He respectively.
This compares well with calculated values of 0.015 and 0.3 from 
Figs.~\ref{fig:be11.overlap} and \ref{fig:he6.overlap}.
These values greatly improve on the estimates given in 
Section~\ref{SEC:VALIDITY}, and give the maximum overlap to the correct order of 
magnitude for scattering systems with core absorption.

\section{Summary \label{SEC:CONCLUSION}}
We have calculated first order non-adiabatic corrections for the first time 
using the core recoil model, in which the valence-target interaction is 
neglected.
Two reactions were studied: the elastic scattering of $^{11}$Be and $^6$He from 
a $^{12}$C target at 10 MeV/nucleon.
The non-adiabatic corrections were compared to previous qualitative estimates of 
the validity of the adiabatic approximation.

The eikonal approximation was used to gain insights into the nature of the 
non-adiabatic corrections.
They were found to be dependent on the overlap of a correction term 
(Eq.~(\ref{chi-tilde})) and the $S$-matrix.
The correction term was found to be strongly dependent on the ratio of the 
valence mass to that of the core, as with the qualitative estimates.
Along with the expected energy dependence, there was also a dependence on the 
strength and diffuseness of the core-target interaction.
The overlap of the correction term with the $S$-matrix was strongly dependent on 
the strength of the imaginary potential for the core-target interaction.
Strong core absorption kills of most of the non-adiabatic corrections as the 
maximum of the correction term lies in a region where the $S$-matrix is zero, 
whilst smaller imaginary potentials increase the overlap of the of the 
$S$-matrix with the correction term producing larger non-adiabatic corrections.
The corrections calculated are much smaller than what is expected from 
qualitative estimates due to the key role that core absorption plays.

An improved estimate for the validity of the adiabatic approximation, when the 
valence-target interaction is neglected, is given in Section~\ref{SEC:NEW-EST}, 
which includes the role of core absorption.
This new estimate recognises that when the core-target absorption is strong, it 
is the size of the correction term in the region near the strong absorption 
radius which is of importance in determining the size of the non-adiabatic 
corrections.
Eq.~(\ref{new-est}) gives a value which is of the correct order of magnitude for 
the size of the non-adiabatic corrections when core absorption is present.

In the core recoil model we have used here, non-adiabatic corrections can only 
arise through projectile excitations occurring from recoil of the core in its 
scattering by the target.
If the valence-target interaction is to included, corrections could also arise 
through recoil of the valence particle.
The strong dependence on the valence-to-core mass ratio is then expected to be 
of great importance in the contribution from these different processes.
The corrections for $^{11}$Be, although being very small partly due to the small 
valence-to-core mass ratio, could be much larger when valence particle recoil is 
included.
This will be dealt with elsewhere \cite{phd-ncs,nac-3bg}.

\begin{acknowledgments}
The authors would like to thank Profs. I.~J.\ Thompson and J.~A.\ Tostevin for 
useful discussions and help in running \textsc{fresco}, and J.A.T. for supplying 
numerical routines.
This work is supported by the United Kingdom Engineering and Physical Sciences 
Research Council (EPSRC) through grant number GR/M82141.
The support of an EPSRC research studentship (for N.C.S.) is also gratefully 
acknowledged.
\end{acknowledgments}

\appendix*

\section{Operator Identity \label{APP:OP}}
We are interested in the operator product of the three factors 
$\psi^{*}(\vec{r})$, ($\hme$), and $\phi(\vec{r})$.
Our result is that this can be re-expressed as
\be\begin{split} \label{id1}
\psi^*(\hme)\phi(\vec{r}) =& -\cona\del{r}\psi^*\phi\cdot\del{r} \\* &{}+ 
\cona[(\del{r}\psi^{*})\phi-\psi^{*}(\del{r}\phi)]\cdot\del{r}. 
\end{split}\ee
The result assumes that $\phi$ is an eigenstate of $H=-\cona\delsq{r}+V$ with 
eigenvalue $\varepsilon$, but $\psi$ can be arbitrary.
$V$ must be a local operator.
$\phi$ and $\psi$ do not have to be bound states.

For an $s$-wave state $\phi $ in a real potential, $\phi$ can be assumed to be 
real and the last term in Eq.~(\ref{id1}) vanishes if $\psi=\phi$:   
\be \label{id2}
\phi(r)(\hme)\phi(r)=-\cona\del{r}(\phi)^2\cdot\del{r}.
\ee

In proving Eq.~(\ref{id1}) we use round brackets to indicate when the 
$\vec\nabla$ operator acts only locally on the functions inside the brackets.
We have
\bea
\del{r}\psi^*\phi\cdot\del{r} &=& 
(\del{r}\psi^*\phi)\cdot\del{r}+(\psi^*\phi)\delsq{r} \nonumber \\*
&=& (\del{r}\psi^*\phi)\cdot\del{r} \nonumber \\*
&&{}+ \psi^*\delsq{r}\phi-\psi^*(\delsq{r}\phi)-2\psi^*(\del{r}\phi)\cdot\del{r} 
\nonumber \\*
&=& [(\del{r}\psi^*)\phi-\psi^*(\del{r}\phi)]\cdot\del{r} \nonumber \\*
&&{}+ \psi^*\delsq{r}\phi-\psi^*(\delsq{r}\phi) \nonumber \\*
&=& [(\del{r}\psi^*)\phi-\psi^*(\del{r}\phi)]\cdot\del{r} \nonumber \\*
&&{}+ \psi^*\delsq{r}\phi-\psi^*\frac{2\mu}{\hbar^2}((V-\varepsilon)\phi), 
\label{p1}
\eea
where we have put double brackets around the last factor to emphasise that $V$ 
acts on $\phi$ only.
We can drop these brackets if $V$ is a local operator.

The last line in Eq.~(\ref{p1}) can now be re-expressed in terms of $H$ and we 
obtain
\bea
\del{r}\psi^*\phi\cdot\del{r} &=& 
[(\del{r}\psi^*)\phi-\psi^*(\del{r}\phi)]\cdot\del{r} \nonumber \\*
&&{}- 
\frac{2\mu}{\hbar^2}\psi^*[-\frac{\hbar^2}{2\mu}\delsq{r}+V-\varepsilon]\phi 
\nonumber \\*
&=& [(\del{r}\psi^*)\phi-\psi^*(\del{r}\phi)]\cdot\del{r} \nonumber \\*
&&{}- \frac{2\mu}{\hbar^2}\psi^*(\hme)\phi. \label{p2}
\eea
This is equivalent to the identity given in Eq.~(\ref{id1}).

The identity (\ref{id2}) is frequently seen in a form which is equivalent to it 
when $\phi(r)$ is nodeless: 
\bea
\hme &=& -\frac{\hbar^2}{2\mu}\phi^{-1}\del{r}\phi^2\cdot\del{r}\phi^{-1} 
\nonumber \\*
&=& \frac{\vec{A}^{(+)}.\vec{A}^{(-)}}{2\mu}. \label{id4}
\eea
where
\bea
\vec{A}^{(+)} &=& \phi^{-1}\vec{p}\phi \\*
\vec{A}^{(-)} &=& \phi\vec{p}\phi^{-1},
\eea
and $\vec{p}=-\im\hbar\del{r}$ is the momentum operator.
The factorised form (\ref{id4}) is used in 1-dimension in the formulation of 
super-symmetric quantum mechanics \cite{Suk1}.

\bibliography{ref}

\begin{thebibliography}{27}
\expandafter\ifx\csname natexlab\endcsname\relax\def\natexlab#1{#1}\fi
\expandafter\ifx\csname bibnamefont\endcsname\relax
  \def\bibnamefont#1{#1}\fi
\expandafter\ifx\csname bibfnamefont\endcsname\relax
  \def\bibfnamefont#1{#1}\fi
\expandafter\ifx\csname citenamefont\endcsname\relax
  \def\citenamefont#1{#1}\fi
\expandafter\ifx\csname url\endcsname\relax
  \def\url#1{\texttt{#1}}\fi
\expandafter\ifx\csname urlprefix\endcsname\relax\def\urlprefix{URL }\fi
\providecommand{\bibinfo}[2]{#2}
\providecommand{\eprint}[2][]{\url{#2}}

\bibitem[{\citenamefont{Johnson and Soper}(1970)}]{adia}
\bibinfo{author}{\bibfnamefont{R.~C.} \bibnamefont{Johnson}} \bibnamefont{and}
  \bibinfo{author}{\bibfnamefont{P.~J.~R.} \bibnamefont{Soper}},
  \bibinfo{journal}{Phys.\ Rev.\ C} \textbf{\bibinfo{volume}{1}},
  \bibinfo{pages}{976} (\bibinfo{year}{1970}).

\bibitem[{\citenamefont{Johnson et~al.}(1997)\citenamefont{Johnson, Al-Khalili,
  and Tostevin}}]{crm-prl}
\bibinfo{author}{\bibfnamefont{R.~C.} \bibnamefont{Johnson}},
  \bibinfo{author}{\bibfnamefont{J.~S.} \bibnamefont{Al-Khalili}},
  \bibnamefont{and} \bibinfo{author}{\bibfnamefont{J.~A.}
  \bibnamefont{Tostevin}}, \bibinfo{journal}{Phys.\ Rev.\ Lett.}
  \textbf{\bibinfo{volume}{79}}, \bibinfo{pages}{2771} (\bibinfo{year}{1997}).

\bibitem[{\citenamefont{Tostevin et~al.}(1998{\natexlab{a}})}]{crm-coul-bu-d}
\bibinfo{author}{\bibfnamefont{J.~A.} \bibnamefont{Tostevin}}
  \bibnamefont{et~al.}, \bibinfo{journal}{Phys.\ Lett.\ B}
  \textbf{\bibinfo{volume}{424}}, \bibinfo{pages}{219}
  (\bibinfo{year}{1998}{\natexlab{a}}).

\bibitem[{\citenamefont{Tostevin
  et~al.}(1998{\natexlab{b}})\citenamefont{Tostevin, Rugmai, and
  Johnson}}]{crm-coul-bu+heavy}
\bibinfo{author}{\bibfnamefont{J.~A.} \bibnamefont{Tostevin}},
  \bibinfo{author}{\bibfnamefont{S.}~\bibnamefont{Rugmai}}, \bibnamefont{and}
  \bibinfo{author}{\bibfnamefont{R.~C.} \bibnamefont{Johnson}},
  \bibinfo{journal}{Phys.\ Rev.\ C} \textbf{\bibinfo{volume}{57}},
  \bibinfo{pages}{3225} (\bibinfo{year}{1998}{\natexlab{b}}).

\bibitem[{\citenamefont{Banerjee
  et~al.}(1998{\natexlab{a}})\citenamefont{Banerjee, Thompson, and
  Tostevin}}]{crm-coul-bu-1n}
\bibinfo{author}{\bibfnamefont{P.}~\bibnamefont{Banerjee}},
  \bibinfo{author}{\bibfnamefont{I.~J.} \bibnamefont{Thompson}},
  \bibnamefont{and} \bibinfo{author}{\bibfnamefont{J.~A.}
  \bibnamefont{Tostevin}}, \bibinfo{journal}{Phys.\ Rev.\ C}
  \textbf{\bibinfo{volume}{58}}, \bibinfo{pages}{1042}
  (\bibinfo{year}{1998}{\natexlab{a}}).

\bibitem[{\citenamefont{Banerjee
  et~al.}(1998{\natexlab{b}})\citenamefont{Banerjee, Tostevin, and
  Thompson}}]{crm-coul-bu-2n}
\bibinfo{author}{\bibfnamefont{P.}~\bibnamefont{Banerjee}},
  \bibinfo{author}{\bibfnamefont{J.~A.} \bibnamefont{Tostevin}},
  \bibnamefont{and} \bibinfo{author}{\bibfnamefont{I.~J.}
  \bibnamefont{Thompson}}, \bibinfo{journal}{Phys.\ Rev.\ C}
  \textbf{\bibinfo{volume}{58}}, \bibinfo{pages}{1337}
  (\bibinfo{year}{1998}{\natexlab{b}}).

\bibitem[{\citenamefont{Yahiro et~al.}(1986)\citenamefont{Yahiro, Iseri,
  Kameyama, Kamimura, and Kawai}}]{deuteron-review}
\bibinfo{author}{\bibfnamefont{M.}~\bibnamefont{Yahiro}},
  \bibinfo{author}{\bibfnamefont{Y.}~\bibnamefont{Iseri}},
  \bibinfo{author}{\bibfnamefont{H.}~\bibnamefont{Kameyama}},
  \bibinfo{author}{\bibfnamefont{M.}~\bibnamefont{Kamimura}}, \bibnamefont{and}
  \bibinfo{author}{\bibfnamefont{M.}~\bibnamefont{Kawai}},
  \bibinfo{journal}{Prog.\ Theo.\ Phys.\ Suppl.} \textbf{\bibinfo{volume}{89}},
  \bibinfo{pages}{32} (\bibinfo{year}{1986}).

\bibitem[{\citenamefont{Johnson}(1999)}]{nac-greece}
\bibinfo{author}{\bibfnamefont{R.~C.} \bibnamefont{Johnson}}, in
  \emph{\bibinfo{booktitle}{Proc.\ Eur.\ Conf.\ On Advances in Nuclear Physics
  and Related Areas, Thessaloniki, Greece, July 8-12, 1997}}, edited by
  \bibinfo{editor}{\bibfnamefont{D.~M.} \bibnamefont{Brink}},
  \bibinfo{editor}{\bibfnamefont{M.~E.} \bibnamefont{Grypeos}},
  \bibnamefont{and} \bibinfo{editor}{\bibfnamefont{S.~E.} \bibnamefont{Massen}}
  (\bibinfo{publisher}{Giahoudi-Giapouli}, \bibinfo{address}{Thessaloniki},
  \bibinfo{year}{1999}), p. \bibinfo{pages}{297}.

\bibitem[{\citenamefont{Glauber}(1959)}]{Glauber}
\bibinfo{author}{\bibfnamefont{R.~J.} \bibnamefont{Glauber}}, in
  \emph{\bibinfo{booktitle}{Lectures in Theoretical Physics}}, edited by
  \bibinfo{editor}{\bibfnamefont{W.~E.} \bibnamefont{Brittin}}
  (\bibinfo{publisher}{Interscience}, \bibinfo{address}{N.Y.},
  \bibinfo{year}{1959}), vol.~\bibinfo{volume}{1}, p. \bibinfo{pages}{315}.

\bibitem[{\citenamefont{Al-Khalili and Johnson}(1992)}]{Gl-d}
\bibinfo{author}{\bibfnamefont{J.~S.} \bibnamefont{Al-Khalili}}
  \bibnamefont{and} \bibinfo{author}{\bibfnamefont{R.~C.}
  \bibnamefont{Johnson}}, \bibinfo{journal}{Nucl.\ Phys.\ A}
  \textbf{\bibinfo{volume}{546}}, \bibinfo{pages}{622} (\bibinfo{year}{1992}).

\bibitem[{\citenamefont{Al-Khalili et~al.}(1995)\citenamefont{Al-Khalili,
  Thompson, and Tostevin}}]{4body-coul}
\bibinfo{author}{\bibfnamefont{J.~S.} \bibnamefont{Al-Khalili}},
  \bibinfo{author}{\bibfnamefont{I.~J.} \bibnamefont{Thompson}},
  \bibnamefont{and} \bibinfo{author}{\bibfnamefont{J.~A.}
  \bibnamefont{Tostevin}}, \bibinfo{journal}{Nucl.\ Phys.\ A}
  \textbf{\bibinfo{volume}{581}}, \bibinfo{pages}{331} (\bibinfo{year}{1995}).

\bibitem[{\citenamefont{Al-Khalili et~al.}(1996)\citenamefont{Al-Khalili,
  Tostevin, and Thompson}}]{halo-radii}
\bibinfo{author}{\bibfnamefont{J.~S.} \bibnamefont{Al-Khalili}},
  \bibinfo{author}{\bibfnamefont{J.~A.} \bibnamefont{Tostevin}},
  \bibnamefont{and} \bibinfo{author}{\bibfnamefont{I.~J.}
  \bibnamefont{Thompson}}, \bibinfo{journal}{Phys.\ Rev.\ C}
  \textbf{\bibinfo{volume}{54}}, \bibinfo{pages}{1843} (\bibinfo{year}{1996}).

\bibitem[{\citenamefont{Al-Khalili and Tostevin}(1996)}]{halo-radii-l}
\bibinfo{author}{\bibfnamefont{J.~S.} \bibnamefont{Al-Khalili}}
  \bibnamefont{and} \bibinfo{author}{\bibfnamefont{J.~A.}
  \bibnamefont{Tostevin}}, \bibinfo{journal}{Phys.\ Rev.\ Lett.}
  \textbf{\bibinfo{volume}{76}}, \bibinfo{pages}{3903} (\bibinfo{year}{1996}).

\bibitem[{\citenamefont{Al-Khalili et~al.}(1997)\citenamefont{Al-Khalili,
  Tostevin, and Brooke}}]{non-eikonal}
\bibinfo{author}{\bibfnamefont{J.~S.} \bibnamefont{Al-Khalili}},
  \bibinfo{author}{\bibfnamefont{J.~A.} \bibnamefont{Tostevin}},
  \bibnamefont{and} \bibinfo{author}{\bibfnamefont{J.~M.}
  \bibnamefont{Brooke}}, \bibinfo{journal}{Phys.\ Rev.\ C}
  \textbf{\bibinfo{volume}{55}}, \bibinfo{pages}{R1018} (\bibinfo{year}{1997}).

\bibitem[{\citenamefont{Brooke et~al.}(1999)\citenamefont{Brooke, Al-Khalili,
  and Tostevin}}]{non-eik-sm}
\bibinfo{author}{\bibfnamefont{J.~M.} \bibnamefont{Brooke}},
  \bibinfo{author}{\bibfnamefont{J.~S.} \bibnamefont{Al-Khalili}},
  \bibnamefont{and} \bibinfo{author}{\bibfnamefont{J.~A.}
  \bibnamefont{Tostevin}}, \bibinfo{journal}{Phys.\ Rev.\ C}
  \textbf{\bibinfo{volume}{59}}, \bibinfo{pages}{1560} (\bibinfo{year}{1999}).

\bibitem[{\citenamefont{Amakawa et~al.}(1979)\citenamefont{Amakawa, Yamaji,
  Mori, and Yazaki}}]{adia-method}
\bibinfo{author}{\bibfnamefont{H.}~\bibnamefont{Amakawa}},
  \bibinfo{author}{\bibfnamefont{S.}~\bibnamefont{Yamaji}},
  \bibinfo{author}{\bibfnamefont{A.}~\bibnamefont{Mori}}, \bibnamefont{and}
  \bibinfo{author}{\bibfnamefont{K.}~\bibnamefont{Yazaki}},
  \bibinfo{journal}{Phys.\ Lett.\ B} \textbf{\bibinfo{volume}{82}},
  \bibinfo{pages}{13} (\bibinfo{year}{1979}).

\bibitem[{\citenamefont{Thompson}()}]{adia-code}
\bibinfo{author}{\bibfnamefont{I.~J.} \bibnamefont{Thompson}},
  \bibinfo{note}{computer program \textsc{adia}, {D}aresbury Laboratory Report
  (1984), unpublished}.

\bibitem[{\citenamefont{Christley et~al.}(1997)\citenamefont{Christley,
  Al-Khalili, Tostevin, and Johnson}}]{4b-adia}
\bibinfo{author}{\bibfnamefont{J.~A.} \bibnamefont{Christley}},
  \bibinfo{author}{\bibfnamefont{J.~S.} \bibnamefont{Al-Khalili}},
  \bibinfo{author}{\bibfnamefont{J.~A.} \bibnamefont{Tostevin}},
  \bibnamefont{and} \bibinfo{author}{\bibfnamefont{R.~C.}
  \bibnamefont{Johnson}}, \bibinfo{journal}{Nucl.\ Phys.\ A}
  \textbf{\bibinfo{volume}{624}}, \bibinfo{pages}{275} (\bibinfo{year}{1997}).

\bibitem[{\citenamefont{Johnson}(1998)}]{nac-jpg}
\bibinfo{author}{\bibfnamefont{R.~C.} \bibnamefont{Johnson}},
  \bibinfo{journal}{J.\ Phys.\ G: Nucl.\ Part.\ Phys.}
  \textbf{\bibinfo{volume}{24}}, \bibinfo{pages}{1583} (\bibinfo{year}{1998}).

\bibitem[{\citenamefont{Summers}(2001)}]{phd-ncs}
\bibinfo{author}{\bibfnamefont{N.~C.} \bibnamefont{Summers}}, Ph.D. thesis,
  \bibinfo{school}{University of Surrey} (\bibinfo{year}{2001}),
  \bibinfo{note}{unpublished}.

\bibitem[{\citenamefont{Summers et~al.}()\citenamefont{Summers, Johnson, and
  Al-Khalili}}]{nac-3bg}
\bibinfo{author}{\bibfnamefont{N.~C.} \bibnamefont{Summers}},
  \bibinfo{author}{\bibfnamefont{R.~C.} \bibnamefont{Johnson}},
  \bibnamefont{and} \bibinfo{author}{\bibfnamefont{J.~S.}
  \bibnamefont{Al-Khalili}}, \bibinfo{note}{to be published}.

\bibitem[{\citenamefont{Gl\"ockle}(2002)}]{glockle}
\bibinfo{author}{\bibfnamefont{W.}~\bibnamefont{Gl\"ockle}}, in
  \emph{\bibinfo{booktitle}{Scattering, Scattering and Inverse Scattering in
  Pure and Applied Science}}, edited by
  \bibinfo{editor}{\bibfnamefont{R.}~\bibnamefont{Pike}} \bibnamefont{and}
  \bibinfo{editor}{\bibfnamefont{P.}~\bibnamefont{Sabatier}}
  (\bibinfo{publisher}{Academic Press}, \bibinfo{address}{London},
  \bibinfo{year}{2002}), vol.~\bibinfo{volume}{2}, chap.
  \bibinfo{chapter}{3.1.2}.

\bibitem[{\citenamefont{Thompson}(1988)}]{fresco}
\bibinfo{author}{\bibfnamefont{I.~J.} \bibnamefont{Thompson}},
  \bibinfo{journal}{Comput.\ Phys.\ Rep.} \textbf{\bibinfo{volume}{7}},
  \bibinfo{pages}{167} (\bibinfo{year}{1988}).

\bibitem[{\citenamefont{Messiah}(1970)}]{messiah1}
\bibinfo{author}{\bibfnamefont{A.}~\bibnamefont{Messiah}},
  \emph{\bibinfo{title}{Quantum Mechanics}}, vol.~\bibinfo{volume}{1}
  (\bibinfo{publisher}{North-Holland}, \bibinfo{address}{Amsterdam},
  \bibinfo{year}{1970}), \bibinfo{note}{translated from the French by G. M.
  Temmer}.

\bibitem[{\citenamefont{Al-Khalili and Tostevin}(2002)}]{jak-jat-scat}
\bibinfo{author}{\bibfnamefont{J.~S.} \bibnamefont{Al-Khalili}}
  \bibnamefont{and} \bibinfo{author}{\bibfnamefont{J.~A.}
  \bibnamefont{Tostevin}}, in \emph{\bibinfo{booktitle}{Scattering, Scattering
  and Inverse Scattering in Pure and Applied Science}}, edited by
  \bibinfo{editor}{\bibfnamefont{R.}~\bibnamefont{Pike}} \bibnamefont{and}
  \bibinfo{editor}{\bibfnamefont{P.}~\bibnamefont{Sabatier}}
  (\bibinfo{publisher}{Academic Press}, \bibinfo{address}{London},
  \bibinfo{year}{2002}), vol.~\bibinfo{volume}{2}, chap.
  \bibinfo{chapter}{3.1.3}.

\bibitem[{\citenamefont{Sakurai}(1994)}]{sakurai}
\bibinfo{author}{\bibfnamefont{J.~J.} \bibnamefont{Sakurai}},
  \emph{\bibinfo{title}{Modern Quantum Mechanics}}
  (\bibinfo{publisher}{Addison-Wesley}, \bibinfo{address}{Reading, Mass.;
  Wokingham}, \bibinfo{year}{1994}), \bibinfo{edition}{revised} ed.

\bibitem[{\citenamefont{Sukumar}(1985)}]{Suk1}
\bibinfo{author}{\bibfnamefont{C.~V.} \bibnamefont{Sukumar}},
  \bibinfo{journal}{J.\ Phys.\ A: Nucl.\ Math.\ Gen.}
  \textbf{\bibinfo{volume}{18}}, \bibinfo{pages}{2917} (\bibinfo{year}{1985}).

\end{thebibliography}

\end{document}